\PassOptionsToPackage{unicode}{hyperref}
\PassOptionsToPackage{hyphens}{url}
\documentclass[
  american,
]{article}
\usepackage{lmodern}
\usepackage{amssymb,amsmath}
\usepackage{ifxetex,ifluatex}
\ifnum 0\ifxetex 1\fi\ifluatex 1\fi=0 % if pdftex
  \usepackage[T1]{fontenc}
  \usepackage[utf8]{inputenc}
  \usepackage{textcomp} % provide euro and other symbols
\else % if luatex or xetex
  \usepackage{unicode-math}
  \defaultfontfeatures{Scale=MatchLowercase}
  \defaultfontfeatures[\rmfamily]{Ligatures=TeX,Scale=1}
\fi
\IfFileExists{upquote.sty}{\usepackage{upquote}}{}
\IfFileExists{microtype.sty}{% use microtype if available
  \usepackage[]{microtype}
  \UseMicrotypeSet[protrusion]{basicmath} % disable protrusion for tt fonts
}{}
\makeatletter
\@ifundefined{KOMAClassName}{% if non-KOMA class
  \IfFileExists{parskip.sty}{%
    \usepackage{parskip}
  }{% else
    \setlength{\parindent}{0pt}
    \setlength{\parskip}{6pt plus 2pt minus 1pt}}
}{% if KOMA class
  \KOMAoptions{parskip=half}}
\makeatother
\usepackage{xcolor}
\IfFileExists{xurl.sty}{\usepackage{xurl}}{} % add URL line breaks if available
\IfFileExists{bookmark.sty}{\usepackage{bookmark}}{\usepackage{hyperref}}
\hypersetup{
  pdftitle={Efficient leave-one-out cross-validation for Bayesian non-factorized normal and Student-t models},
  pdfauthor={Paul-Christian Bürkner \^{}\{1*\}, Jonah Gabry \^{}2, \& Aki Vehtari \^{}1},
  pdflang={en-US},
  hidelinks,
  pdfcreator={LaTeX via pandoc}}
\usepackage[margin=1in]{geometry}
\usepackage{longtable,booktabs}
\usepackage{etoolbox}
\makeatletter
\patchcmd\longtable{\par}{\if@noskipsec\mbox{}\fi\par}{}{}
\makeatother
\IfFileExists{footnotehyper.sty}{\usepackage{footnotehyper}}{\usepackage{footnote}}
\makesavenoteenv{longtable}
\usepackage{graphicx,grffile}
\makeatletter
\def\maxwidth{\ifdim\Gin@nat@width>\linewidth\linewidth\else\Gin@nat@width\fi}
\def\maxheight{\ifdim\Gin@nat@height>\textheight\textheight\else\Gin@nat@height\fi}
\makeatother
\setkeys{Gin}{width=\maxwidth,height=\maxheight,keepaspectratio}
\makeatletter
\def\fps@figure{htbp}
\makeatother
\usepackage{amsmath}
\usepackage[utf8]{inputenc}
\usepackage[T1]{fontenc}
\usepackage{setspace}
\newtheorem{proposition}{Proposition}

\usepackage{booktabs}
\usepackage{longtable}
\usepackage{array}
\usepackage{multirow}
\usepackage{wrapfig}
\usepackage{float}
\usepackage{colortbl}
\usepackage{pdflscape}
\usepackage{tabu}
\usepackage{threeparttable}
\usepackage{threeparttablex}
\usepackage[normalem]{ulem}
\usepackage{makecell}
\usepackage{xcolor}
\ifxetex
  \usepackage{polyglossia}
  \setmainlanguage[variant=american]{english}
\else
  \usepackage[shorthands=off,main=american]{babel}
\fi
\usepackage[]{natbib}
\title{Efficient leave-one-out cross-validation for Bayesian non-factorized normal and Student-\(t\) models}
\author{Paul-Christian Bürkner \(^{1*}\), Jonah Gabry \(^2\), \& Aki Vehtari \(^1\)}
\date{\(^1\) Department of Computer Science, Aalto University, Finland\break
\(^2\) Applied Statistics Center and ISERP, Columbia University, USA \break
\(^*\) Corresponding author, Email: \href{mailto:paul.buerkner@gmail.com}{\nolinkurl{paul.buerkner@gmail.com}}}

\begin{document}
\maketitle
\begin{abstract}
Cross-validation can be used to measure a model's predictive accuracy for the
purpose of model comparison, averaging, or selection. Standard leave-one-out
cross-validation (LOO-CV) requires that the observation model can be
factorized into simple terms, but a lot of important models in temporal and
spatial statistics do not have this property or are inefficient or unstable
when forced into a factorized form. We derive how to efficiently compute and
validate both exact and approximate LOO-CV for any Bayesian non-factorized
model with a multivariate normal or Student-\(t\) distribution on the outcome
values. We demonstrate the method using lagged simultaneously autoregressive
(SAR) models as a case study.
\linebreak\linebreak 
\end{abstract}

\hypertarget{introduction}{%
\section{Introduction}\label{introduction}}

In the absence of new data, cross-validation is a general approach for
evaluating a statistical model's predictive accuracy for the purpose of model
comparison, averaging, or selection \citep{geisser1979, hoeting1999, ando2010, vehtari2012}. One widely used variant of cross-validation is
\emph{leave-one-out cross-validation} (LOO-CV), where observations are left out
one at a time and then predicted based on the model fit to the remaining data.
Predictive accuracy is evaluated by first computing a pointwise predictive
measure and then taking the sum of these values over all observations to obtain
a single measure of predictive accuracy \citep[e.g.,][]{vehtari2017loo}.
In this paper, we focus on the expected log predictive density (ELPD) as the measure
of predictive accuracy. The ELPD takes the whole predictive distribution into
account and is less focused on the bulk of the distribution compared to other common
measures such as the root mean squared error (RMSE) or mean absolute error \citep[MAE; see][ for details]{vehtari2012}. Exact LOO-CV is costly, as it requires fitting the
model as many times as there are observations in the data. Depending on the size
of the data, complexity of the model, and estimation method, this can be
practically infeasible as it simply requires too much computation time. For this
reason, fast approximate versions of LOO-CV have been developed \citep[
\citet{vehtari2017loo}]{gelfand1992}, most recently using Pareto-smoothed importance-sampling \citep[PSIS;][]{vehtari2017loo, vehtari2019psis}.

A standard assumption of any such fast LOO-CV approach using the ELPD is that
the model over all observations has to have a factorized form. That
is, the overall observation model should be represented as the product of the
pointwise models for each observation. However, many important
models do not have this factorization property. Particularly in temporal and
spatial statistics it is common to fit multivariate normal or
Student-\(t\) models that have structured covariance matrices such that the model
does not factorize. This is typically due to the fact that observations depend
on other observations from different time periods or different spatial units in
addition to the dependence on the model parameters. Some of these models are
actually non-factorizable, that is, we do not know of any reformulation that
converts the observation model into a factorized form. Other non-factorized
models could be factorized in theory but it is sometimes more robust or
efficient to marginalize out certain parameters, for instance
observation-specific latent variables, and then work with a non-factorized model
instead.

Conceptually, a factorized model is not required for cross-validation in
general or LOO-CV in particular to make sense. This also implies that neither
conditional independence nor conditional exchangability are necessary
assumptions. However, when using non-factorized observation models
in LOO-CV, computational challenges arise.
In this paper, we derive how to perform efficient approximate LOO-CV for
\emph{any} Bayesian multivariate normal or Student-\(t\) model with an invertible
covariance or scale matrix, regardless of whether or not the model
factorizes. We also provide equations for computing exact LOO-CV for these
models, which can be used to validate the approximation and to handle
problematic observations. Throughout, a Bayesian model specification
and estimation via Markov chain Monte Carlo (MCMC) is assumed. We have
implemented the developed methods in the R package brms \citep{brms1, brms2}.
All materials including R source code are available in an online
supplement\footnote{Supplemental materials available at
\url{https://github.com/paul-buerkner/psis-non-factorized-paper}.}.

\hypertarget{pwnf}{%
\section{Pointwise log-likelihood for non-factorized models}\label{pwnf}}

When computing ELPD-based \emph{exact} LOO-CV for a Bayesian model we need to
compute the log leave-one-out predictive densities \(\log{p(y_i | y_{-i})}\) for
every response value \(y_i, \: i = 1, \ldots, N\), where \(y_{-i}\) denotes all
response values except observation \(i\). To obtain \(p(y_i | y_{-i})\), we need to
have access to the pointwise likelihood \(p(y_i\,|\, y_{-i}, \theta)\) and
integrate over the model parameters \(\theta\):
\begin{equation}
\label{loo-pd}
p(y_i\,|\,y_{-i}) =
  \int p(y_i\,|\, y_{-i}, \theta) \, p(\theta\,|\, y_{-i}) \,d \theta
\end{equation}
Here, \(p(\theta\,|\, y_{-i})\) is the leave-one-out posterior distribution for
\(\theta\), that is, the posterior distribution for \(\theta\) obtained by fitting
the model while holding out the \(i\)th observation (in Section
\ref{approx-loo-cv}, we will show how refitting the model to data \(y_{-i}\) can
be avoided).

\begin{figure}
\centering
\begin{minipage}{.48\textwidth}
  \centering
  \includegraphics{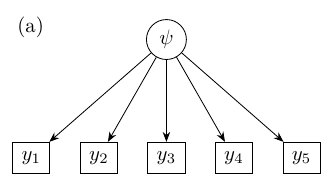}
  %\captionof{figure}{A figure}
\end{minipage}
\begin{minipage}{.48\textwidth}
  \centering
  \includegraphics{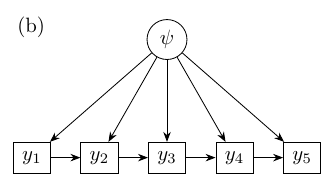}
  %\captionof{figure}{Another figure}
\end{minipage}
\begin{minipage}{.48\textwidth}
  \centering
  \includegraphics{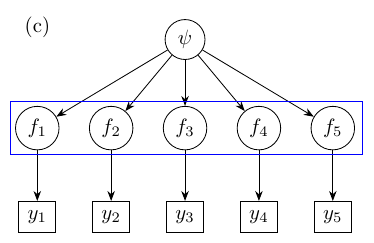}
  %\captionof{figure}{A figure}
\end{minipage}
\begin{minipage}{.48\textwidth}
  \centering
  \includegraphics{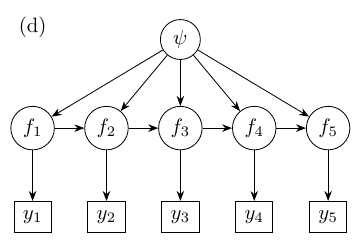}
  %\captionof{figure}{Another figure}
\end{minipage}
\begin{minipage}{.48\textwidth}
  \centering
  \includegraphics{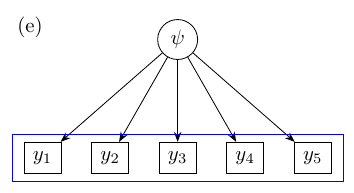}
  %\captionof{figure}{Another figure}
\end{minipage}
\caption{Directional graphs illustrating common observation model dependency structures
schematically. Black rectangles depict manifest variables, that is, observed
response values. Black circles depict latent variables and parameters. Blue
rectangles indicate a joint prior over the surrounded variables. (a)
Conditionally independent responses given hyperparameters (e.g., a linear
regression model). (b) Conditionally dependent responses with a Markovian property
(e.g., an autoregressive model of order 1). (c) Conditionally independent
responses given observation-specific latent variables with a joint prior (e.g.,
a latent Gaussian process model). (d)  Conditionally independent responses given
observation-specific latent variables with a Markov property (e.g., a hidden
Markov model). (e) Non-factorized model with a joint observation model over all
responses.}
\label{fig:dags}
\end{figure}

If the observation model is formulated directly as the product of the
pointwise observation models, we call it a \emph{factorized}
model. In this case, the likelihood is also the product of the pointwise
likelihood contributions \(p(y_i\,|\, y_{-i}, \theta)\).
To better illustrate possible structures of the observation models, we
formally divide \(\theta\) into two parts, observation-specific latent
variables \(f = (f_1, \ldots, f_N)\) and hyperparameters \(\psi\), so that
\(p(y_i\,|\, y_{-i}, \theta) = p(y_i\,|\, y_{-i}, f_i, \psi)\).
Depending on the model, one of the two parts of \(\theta\)
may also be empty. In very simple models, such as linear regression models,
latent variables are not explicitely presented and response values are
conditionally independent given
\(\psi\), so that \(p(y_i\,|\, y_{-i}, f_i, \psi) = p(y_i \,|\, \psi)\) (see Figure
\ref{fig:dags} (a)). The full likelihood can then be written in the familiar
form
\begin{equation}
p(y \,|\, \psi) = \prod_{i=1}^N p(y_i \,|\, \psi),
\end{equation}
where \(y = (y_1, \ldots, y_N)\) denotes the vector of all responses.
When the likelihood factorizes this way, the conditional
pointwise log-likelihood can be obtained easily by computing
\(p(y_i\,|\, \psi)\) for each \(i\) with computational cost \(O(n)\).

If directional paths between consecutive responses are added,
responses are no longer conditionally independent, but the model
factorizes to simple terms with Markovian dependency. This is common
in time-series models. For example, in an
autoregressive model of order 1 (see Figure \ref{fig:dags} (b)), the pointwise
likelihoods are given by \(p(y_i \,|\, y_{i-1}, \psi)\). In other time series, models
may have observation-specific latent variables \(f_i\) and conditionally independent
responses so that the pointwise log-likelihoods simplify to
\(p(y_i\,|\, y_{-i}, f_i, \psi) = p(y_i \,|\, f_i)\).
In models without directional paths between the
latent values \(f\) (see Figure \ref{fig:dags} (c)), such as latent Gaussian
processes \citep[GPs; e.g.,][]{rasmussen2003} or spatial conditional autoregressive
(CAR) models \citep[e.g.,][]{gelfand2003}, an explicit joint prior over \(f\) is imposed.
In models with directional paths between the latent values \(f\) (see Figure
\ref{fig:dags} (d)), such as hidden Markov models \citep[HMMs; e.g.,][]{rabiner1986}, the
joint prior over \(f\) is defined implicitly via the directional dependencies. What is
more, estimation can make use of the latent Markov property of such models, for
example, using the Kalman filter \citep[e.g.,][]{welch1995}. In all of these cases
(i.e., Figure \ref{fig:dags} (a) to (d)), the factorization property is retained and
computational cost for the pointwise log-likelihood contributions remains in
\(O(n)\).

Yet, there are several reasons why a \emph{non-factorized} observation model
(see Figure \ref{fig:dags} (e)) may be necessary or preferred. In non-factorized
models, the joint likelihood of the response values \(p(y \,|\, \theta)\) is not
factorized into observation-specific components, but rather given directly as one
joint expression. For some models, an analytical factorized formulation is
simply not available in which case we speak of a \emph{non-factorizable} model. Even
in models whose observation model can be factorized in principle, it may still be
preferable to use a non-factorized form. This is true in particular for
models with observation-specific latent variables (see Figure \ref{fig:dags} (c)
and (d)), as a non-factorized formulation where the latent variables
have been integrated out is often more efficient and numerically
stable. For example, a latent GP combined with a Gaussian
observation model can be fit more efficiently by marginalizing over \(f\) and
formulating the GP directly on the responses \(y\) \citep[e.g.,][]{rasmussen2003}.
Such marginalization has the additional advantage that both exact and
approximate leave-one-out predictive estimation become more stable. This is
because, in the factorized formulation, leaving out response \(y_i\) also implies
treating the corresponding latent variable \(f_i\) as missing, which is then only
identified through the joint prior over \(f\). If this prior is weak, the
posterior of \(f_i\) is highly influenced by one observation and the leave-one-out
predictions of \(y_i\) may be unstable both numerically and because of estimation
error due to finite MCMC sampling or similar finite approximations.

Whether a non-factorized model is used by necessity or for efficiency and
stability, it comes at the cost of having no direct access to the leave-one-out
predictive densities \eqref{loo-pd} and thus to the overall leave-one-out
predictive accuracy. In theory, we can express the observation-specific
likelihoods in terms of the joint likelihood via
\begin{equation}
\label{pw-lh}
p(y_i \,|\, y_{i-1}, \theta) = 
  \frac{p(y \,|\, \theta)}{p(y_{-i} \,|\, \theta)} = 
  \frac{p(y \,|\, \theta)}{\int p(y \,|\, \theta) \, d y_i},
\end{equation}
but the expression on the right-hand side of \eqref{pw-lh} may not always
have an analytical solution. Computing \(\log p(y_i \,|\, y_{-i}, \theta)\) for non-factorized models is therefore often impossible, or at least
inefficient and numerically unstable. However, there is a large
class of multivariate normal and Student-\(t\) models for which we will provide
efficient analytical solutions in this paper.

\hypertarget{non-factorized-normal-models}{%
\subsection{Non-factorized normal models}\label{non-factorized-normal-models}}

The density of the \(N\) dimensional multivariate normal distribution
of vector \(y\) is given by
\begin{equation}
\label{mvnormal}
  p(y | \mu, \Sigma) = \frac{1}{\sqrt{(2 \pi)^N |\Sigma|}} 
  \exp \left(-\frac{1}{2}(y - \mu)^{\rm T} \Sigma^{-1} (y - \mu) \right)
\end{equation}
with mean vector \(\mu\) and covariance matrix \(\Sigma\). Often \(\mu\) and \(\Sigma\)
are functions of the model parameters \(\theta\), that is, \(\mu = \mu(\theta)\) and
\(\Sigma = \Sigma(\theta)\), but for notational convenience we omit the potential
dependence of \(\mu\) and \(\Sigma\) on \(\theta\) unless it is relevant. Using
standard multivariate normal theory \citep[e.g.,][]{tong2012}, we know that for the
\(i\)th observation the conditional distribution \(p(y_i | y_{-i}, \theta)\) is
univariate normal with mean
\begin{equation}
\label{cmean}
  \tilde{\mu}_{i} = \mu_i + \sigma_{i,-i} \Sigma^{-1}_{-i} (y_{-i} - \mu_{-i})
\end{equation}
and variance
\begin{equation}
\label{csd}
  \tilde{\sigma}_{i} = \sigma_{ii} + \sigma_{i,-i} \Sigma^{-1}_{-i} \sigma_{-i,i} .
\end{equation}
In the equations above, \(\mu_{-i}\) is the mean vector without the \(i\)th element,
\(\Sigma_{-i}\) is the covariance matrix without the \(i\)th row and column
(\(\Sigma^{-1}_{-i}\) is its inverse), \(\sigma_{i,-i}\) and \(\sigma_{-i,i}\) are the
\(i\)th row and column vectors of \(\Sigma\) without the \(i\)th element, and
\(\sigma_{ii}\) is the \(i\)th diagonal element of \(\Sigma\). Equations
\eqref{cmean} and \eqref{csd} can be used to compute the pointwise
log-likelihood values as
\begin{equation}
\label{lp-normal}
  \log p(y_i \,|\, y_{-i},\theta)
  = - \frac{1}{2}\log(2\pi \tilde{\sigma}_{i}) 
- \frac{1}{2}\frac{(y_i-\tilde{\mu}_{i})^2}{\tilde{\sigma}_{i}}.
\end{equation}
Evaluating equation \eqref{lp-normal} for each \(y_i\) and each posterior draw
\(\theta_s\) then constitutes the input for the LOO-CV computations. However, the
resulting procedure is quite inefficient. Computation is usually dominated by
the \(O(N^k)\) cost of computing \(\Sigma_{-i}^{-1}\), where \(k\) depends on the
structure of \(\Sigma\). If \(\Sigma\) is dense then \(k = 3\). For sparse \(\Sigma\) or
reduced rank computations we have \(2 < k < 3\). And since \(\Sigma_{-i}^{-1}\)
must be computed for each \(i\), the overall complexity is actually
\(O(N^{k + 1})\).

Additionally, if \(\Sigma_{-i}\) also depends on the model parameters \(\theta\) in
a non-trivial manner, which is the case for most models of practical relevance,
then it needs to be inverted for each of the \(S\) posterior draws. Therefore, in
most applications the overall complexity will actually be \(O(S N^{k+1})\), which
will be impractical in most situations. Accordingly, we seek to find
more efficient expressions for \(\tilde{\mu}_{i}\) and \(\tilde{\sigma}_{i}\)
that make these computations feasible in practice.

\begin{proposition}
\label{prop-eff-normal}
If $y$ is multivariate normal with mean vector $\mu$ and covariance matrix
$\Sigma$, then the conditional mean and standard deviation of $y_i$ given $y_{-i}$
for any observation $i$ can be computed as
\begin{equation}
\label{cmean2}
  \tilde{\mu}_{i} = y_i - \frac{g_i}{\bar{\sigma}_{ii}},
\end{equation}
\begin{equation}
\label{csd2}
  \tilde{\sigma}_{i} = \frac{1}{\bar{\sigma}_{ii}},
\end{equation}
where $g_i = \left[\Sigma^{-1} (y - \mu)\right]_i$ and 
$\bar{\sigma}_{ii} = \left[\Sigma^{-1}\right]_{ii}$. 
\end{proposition}

The proof is based on results from \citet{sundararajan2001} and is provided in
the Appendix. Contrary to the brute force
computations in \eqref{cmean} and \eqref{csd}, where \(\Sigma_{-i}\) has to be
inverted separately for each \(i\), equations \eqref{cmean2} and \eqref{csd2} require
inverting the full covariance matrix \(\Sigma\) only once and it can be reused for each
\(i\). This reduces the computational cost to \(O(N^k)\) if \(\Sigma\) is independent
of \(\theta\) and \(O(S N^k)\) otherwise. If the model is parameterized in terms of
the covariance matrix \(\Sigma = \Sigma(\theta)\), it is not possible to reduce
the complexity further as inverting \(\Sigma\) is unavoidable. However, if the
model is parameterized directly through the inverse of \(\Sigma\), that is
\(\Sigma^{-1} = \Sigma^{-1}(\theta)\), the complexity goes down to \(O(S N^2)\).
Note that the latter is not possible in the brute force approach as both
\(\Sigma\) and \(\Sigma^{-1}\) are required.

\hypertarget{non-factorized-student-t-models}{%
\subsection{\texorpdfstring{Non-factorized Student-\(t\) models}{Non-factorized Student-t models}}\label{non-factorized-student-t-models}}

Several generalizations of the multivariate normal distribution have been
suggested, perhaps most notably the multivariate Student-\(t\) distribution
\citep{zellner1976}, which has an additional positive \emph{degrees of freedom} parameter
\(\nu\) that controls the tails of the distribution. If \(\nu\) is small, the tails
are much fatter than those of the normal distribution. If \(\nu\) is large, the
multivariate Student-\(t\) distribution becomes more similar to the corresponding
multivariate normal distribution and is equal to the latter for \(\nu \rightarrow \infty\). As \(\nu\) can be estimated alongside the other model parameters in
Student-\(t\) models, the thickness of the tails is flexibly adjusted based on
information from the observed response values and the prior. The (multivariate)
Student-\(t\) distribution has been studied in various places \citep[e.g.,][]{zellner1976, ohagan1979, fernandez1999, zhang2010, piche2012, shah2014}. For example, Student-\(t\)
processes which are based on the multivariate Student-\(t\) distribution constitute
a generalization of Gaussian processes while retaining most of the latter's
favorable properties \citep{shah2014}\footnote{A Student-\(t\) process is not to be confused with a factorized
  univariate Student-\(t\) likelihood in combination with a Gaussian process on the
  corresponding latent variables as these models have different properties.}.

The density of the \(N\) dimensional multivariate Student-\(t\) distribution
of vector \(y\) is given by
\begin{equation}
\label{mvstudent}
  p(y | \nu, \mu, \Sigma) = \frac{\Gamma((\nu + N) / 2)}{\Gamma(\nu / 2)}
  \frac{1}{\sqrt{(\nu \pi)^N |\Sigma|}} 
  \left(1 + \frac{1}{\nu} (y - \mu)^{\rm T} \Sigma^{-1} (y - \mu) \right)^{-(\nu + N)/2}
\end{equation}
with degrees of freedom \(\nu\), location vector \(\mu\) and scale matrix \(\Sigma\).
The mean of \(y\) is \(\mu\) if \(\nu > 1\) and \(\frac{\nu}{\nu-2}\Sigma\) is the
covariance matrix if \(\nu > 2\). Similar to the multivariate normal case, the
conditional distribution of the \(i\)th observation given all other observations
and the model parameters, \(p(y_i | y_{-i}, \theta)\), can be computed analytically.

\begin{proposition}
\label{prop-cond-student}
If $y$ is multivariate Student-$t$ with degrees of freedom $\nu$, location vector
$\mu$, and scale matrix $\Sigma$, then the conditional distribution of $y_i$
given $y_{-i}$ for any observation $i$ is univariate Student-$t$ with parameters
\begin{equation}
\label{cnu}
\tilde{\nu}_i = \nu + N - 1,
\end{equation}
\begin{equation}
\label{cmeant}
  \tilde{\mu}_{i} = \mu_i + \sigma_{i,-i} \Sigma^{-1}_{-i}(y_{-i} - \mu_{-i}),
\end{equation}
\begin{equation}
\label{csdt}
  \tilde{\sigma}_{i} = \frac{\nu + \beta_{-i}}{\nu + N - 1} 
  \left( \sigma_{ii} + \sigma_{i,-i} \Sigma^{-1}_{-i} \sigma_{-i,i} \right),
\end{equation}
where 
\begin{equation}
\label{cbeta}
\beta_{-i} = (y_{-i} - \mu_{-i})^{\rm T} \Sigma^{-1}_{-i} (y_{-i} - \mu_{-i}).
\end{equation}
\end{proposition}

A proof based on results of \citet{shah2014} is given in the Appendix.
Here \(\tilde{\mu}_{i}\) is the same as in the normal case and \(\tilde{\sigma}_{i}\)
is the same up to the correction factor \(\frac{\nu + \beta_{-i}}{\nu + N - 1}\),
which approaches \(1\) for \(\nu \rightarrow \infty\) as one would expect.
Based on the above equations, we can compute the pointwise log-likelihood
values in the Student-\(t\) case as
\begin{align}
\label{lp-student}
  \log p(y_i \,|\, y_{-i},\theta)
  &= \log (\Gamma((\tilde{\nu}_i + 1) / 2)) - \log (\Gamma(\tilde{\nu}_i / 2))
- \frac{1}{2}\log(\tilde{\nu}_i \pi \tilde{\sigma}_{i} ) \nonumber \\
&\quad - \frac{\tilde{\nu}_i + 1}{2} \log \left(1 + \frac{1}{\tilde{\nu}_i} \frac{(y_i-\tilde{\mu}_{i})^2}{\tilde{\sigma}_{i}} \right).
\end{align}
This approach has the same overall computational cost of \(O(S N^{k+1})\) as
the non-optimized normal case and is therefore quite inefficient.
Fortunately, the efficiency can again be improved.

\begin{proposition}
\label{prop-eff-student}
If $y$ is multivariate Student-$t$ with degrees of freedom $\nu$, location vector
$\mu$, and scale matrix $\Sigma$, then the conditional location and scale of $y_i$
given $y_{-i}$ for any observation $i$ can be computed as
\begin{equation}
\label{cmeant2}
  \tilde{\mu}_{i} = y_i - \frac{g_i}{\bar{\sigma}_{ii}},
\end{equation}
\begin{equation}
\label{csdt2}
  \tilde{\sigma}_{i} = 
  \frac{\nu + \beta_{-i}}{\nu + N - 1} \frac{1}{\bar{\sigma}_{ii}},
\end{equation}
with
\begin{equation}
\label{cbeta2}
  \beta_{-i} = (y_{-i} - \mu_{-i})^{\rm T} \left( \Sigma^{-1} - \frac{\bar{\sigma}_{-i,i} \bar{\sigma}_{-i,i}^{\rm T}}{\bar{\sigma}_{ii}} \right) (y_{-i} - \mu_{-i}),
\end{equation}
where $g_i = \left[\Sigma^{-1} (y - \mu)\right]_i$,
$\bar{\sigma}_{ii} = \left[\Sigma^{-1}\right]_{ii}$, and 
$\bar{\sigma}_{-i,i} = \left[\Sigma^{-1}\right]_{-i,i}$ is the $i$th column vector of $\Sigma^{-1}$ without the $i$th element. 
\end{proposition}

The proof is provided in the Appendix. After inverting \(\Sigma\), computing
\(\beta_{-i}\) for a single \(i\) is an \(O(N^2)\) operation, which needs to be
repeated for each observation. So the cost of computing \(\beta_{-i}\) for all
observations is \(O(N^3)\). The cost of inverting \(\Sigma\) continues to be
\(O(N^k)\) and so the overall cost is dominated by \(O(N^3)\), or \(O(S N^3)\) if
\(\Sigma\) depends on the model parameters \(\theta\) in a non-trivial manner.
Unlike the normal case, we cannot reduce the computational costs below \(O(S N^3)\) even if the model is parameterized directly in terms of \(\Sigma^{-1} = \Sigma^{-1}(\theta)\) and so avoids matrix inversion altogether. However, this is
still substantially more efficient than the brute force approach, which requires
\(O(S N^{k+1}) > O(SN^3)\) operations.

\hypertarget{example-lagged-sar-models}{%
\subsection{Example: Lagged SAR models}\label{example-lagged-sar-models}}

It often requires additional work to take a given multivariate normal or
Student-\(t\) model and express it in the form required to apply the equations for
the predictive mean and standard deviation. Consider, for example, the lagged
simultaneous autoregressive (SAR) model \citep{cressie1992, haining2003, lesage2009}, a spatial model with many applications in both the social sciences
(e.g., economics) and natural sciences (e.g., ecology). The model is given by
\begin{equation}
y = \rho W y + \eta + \epsilon,
\end{equation}
or equivalently
\begin{equation}
(I - \rho W) y = \eta + \epsilon,
\end{equation}
where \(\rho\) is a scalar spatial correlation parameter and \(W\) is a user-defined
matrix of weights. The matrix \(W\) has entries \(w_{ii} = 0\) along the diagonal
and the off-diagonal entries \(w_{ij}\) are larger when units \(i\) and \(j\) are
closer to each other but mostly zero as well. In a linear model, the predictor
term is \(\eta = X \beta\), with design matrix \(X\) and regression coefficients
\(\beta\), but the definition of the lagged SAR model holds for arbitrary \(\eta\),
so these results are not restricted to the linear case. See \citet{lesage2009}, Section
2.3, for a more detailed introduction to SAR models. A general discussion about
predictions of SAR models from a frequentist perspective can be found in
\citet{goulard2017}.

If we have \(\epsilon \sim \mathrm{N}(0, \sigma^2 I)\), with residual variance \(\sigma^2\)
and identity matrix \(I\) of dimension \(N\), it follows that
\begin{equation}
\label{lagsar}
(I - \rho W) y \sim \mathrm{N}(\eta, \sigma^2 I),
\end{equation}
but this standard way of expressing the model is not compatible with
the requirements of Proposition \ref{prop-eff-normal}. To make the lagged SAR model reconcilable with this proposition we need to rewrite it as follows (conditional on
\(\rho\), \(\eta\), and \(\sigma^2\)):
\begin{equation}
y \sim \mathrm{N}\left((I - \rho W)^{-1} \eta,\, \sigma^2 (I - \rho W)^{-1} (I - \rho W)^{-{\rm T}} \right),
\end{equation}
or more compactly, with \(\widetilde{W} = (I - \rho W)\),
\begin{equation}
y \sim \mathrm{N}\left(\widetilde{W}^{-1} \eta,\, \sigma^2  (\widetilde{W}^{\rm T} \widetilde{W})^{-1} \right).
\end{equation}
Written in this way, the lagged SAR model has the required form \eqref{mvnormal}.
Accordingly, we can compute the leave-one-out
predictive densities with Equations \eqref{cmean2} and \eqref{csd2}, replacing
\(\mu\) with \(\widetilde{W}^{-1} \eta\) and taking the covariance matrix \(\Sigma\) to
be \(\sigma^2 (\widetilde{W}^T \widetilde{W})^{-1}\). This implies
\(\Sigma^{-1}=\sigma^{-2}\widetilde{W} \widetilde{W}^T\) and so that the overall
computational cost is dominated by \(\widetilde{W}^{-1} \eta\). In SAR models, \(W\)
is usually sparse and so is \(\widetilde{W}\). Thus, if sparse matrix operations
are used, then the computational cost for \(\Sigma^{-1}\) will be less than \(O(N^2)\)
and for \(\widetilde{W}^{-1}\) it will be less than \(O(N^3)\) depending on number
of non-zeros and the fill pattern. Since \(\widetilde{W}\) depends on the parameter
\(\rho\) in a non-trivial manner, \(\widetilde{W}^{-1}\) needs to be computed for
each posterior draw, which implies an overall computational cost of less than
\(O(S N^3)\).

If the residuals are Student-\(t\) distributed, we can apply analogous transformations as above to arrive at the Student-\(t\) distribution for the responses
\begin{equation}
y \sim \mathrm{t}\left(\nu,\, \widetilde{W}^{-1} \eta,\, \sigma^2  (\widetilde{W}^{\rm T} \widetilde{W})^{-1} \right),
\end{equation}
with computational cost dominated by the computation of the \(\beta_i\)
from Equation \eqref{cbeta2} which is in \(O(S N^3)\).

Studying leave-one-out predictive densities in SAR models is related to
considering impact measures, that is, measures to quantify how changes in the
predicting variables of a given observation \(i\) affect the responses in other
observations \(j \neq i\) as well as the obtained parameter estimates \citep[see][Section 2.7]{lesage2009}. A detailed discussion of this topic it out of scope
of the present paper.

\hypertarget{approx-loo-cv}{%
\section{Approximate LOO-CV for non-factorized models}\label{approx-loo-cv}}

Exact LOO-CV, requires refitting the model \(N\)
times, each time leaving out one observation. Alternatively, it is possible to
obtain an \emph{approximate} LOO-CV using only a single model fit by instead calculating
the pointwise log-predictive density \eqref{loo-pd},
without leaving out any observations, and then applying an
importance sampling correction \citep{gelfand1992}, for example, using
Pareto smoothed importance sampling \citep[PSIS;][]{vehtari2017loo}.

The conditional pointwise log-likelihood matrix of dimension \(S \times N\) is the
only required input to the approximate LOO-CV algorithm from \citet{vehtari2017loo} and
thus the equations provided in Section \ref{pwnf} allow for approximate LOO-CV
for \emph{any} model that can be expressed conditionally in terms of a
multivariate or Student-\(t\) model with invertible covariance/scale matrix
\(\Sigma\); including those where the likelihood does not factorize.

Suppose we have obtained \(S\) posterior draws \(\theta^{(s)}\) \((s=1,\ldots,S)\), from the
\emph{full} posterior \(p(\theta\,|\, y)\) using MCMC or another sampling algorithm.
Then, the pointwise log-predictive density \eqref{loo-pd} can be approximated
as:
\begin{equation}
 p(y_i\,|\, y_{-i}) \approx
   \frac{ \sum_{s=1}^S p(y_i\,|\,y_{-i},\,\theta^{(s)}) \,w_i^{(s)}}{ \sum_{s=1}^S w_i^{(s)}},
\end{equation}
where \(w_i^{(s)}\) are importance weights to be computed
in two steps. First, we obtain the raw importance ratios
\begin{equation}
  r_i^{(s)} \propto \frac{1}{p(y_i \,|\, y_{-i}, \, \theta^{(s)})},
\end{equation}
and then stabilize them using Pareto-smoothed importance-sampling to
obtain the weights \(w_i^{(s)}\) \citep{vehtari2017loo, vehtari2019psis}. The
resulting approximation is referred to as PSIS-LOO-CV \citep{vehtari2017loo}.

For Bayesian models fit using MCMC, the whole procedure of evaluating
and comparing model fit via PSIS-LOO-CV can be summarized as follows:

\begin{enumerate}
\item Fit the model using MCMC obtaining $S$ samples from the posterior
distribution of the parameters $\theta$.
\item For each of the $S$ draws of $\theta$, compute the pointwise
log-likelihood value for each of the $N$ observations in $y$ 
as described in Section \ref{pwnf}. The results can be stored in an $S \times N$ matrix.
\item Run the PSIS algorithm from \cite{vehtari2017loo} on the $S \times N$
matrix obtained in step 2 to obtain a PSIS-LOO-CV estimate.
For convenience, the \texttt{loo} R package
\citep{loo2018} provides this functionality.
\item Repeat the steps 1 to 3 for each model under consideration and
perform model comparison based on the obtained PSIS-LOO-CV estimates.
\end{enumerate}

In the Section \ref{case-study}, we demonstrate this method by performing
approximate LOO-CV for lagged SAR models fit to spatially correlated
crime data.

\hypertarget{exact-loo-cv}{%
\subsection{Validation using exact LOO-CV}\label{exact-loo-cv}}

In order to validate the approximate LOO-CV procedure, and also in order to
allow exact computations to be made for a small number of leave-one-out folds
for which the Pareto-\(k\) diagnostic \citep{vehtari2019psis} indicates an
unstable approximation, we need to consider how we might do \emph{exact}
LOO-CV for a non-factorized model. Here we will provide the necessary
equations and in the supplementary materials we provide code for comparing the
exact and approximate versions.

In the case of those multivariate normal or Student-\(t\) models that have the marginalization property, exact
LOO-CV is relatively straightforward: when refitting the model we can simply
drop the one row and column of the covariance matrix \(\Sigma\) corresponding to the
held out observation without altering the prior of the other observations.
But this does not hold in general for all multivariate normal or Student-\(t\) models
(in particular it does not hold for SAR models).
Instead, in order to keep the original prior, we may need to maintain the full
covariance matrix \(\Sigma\) even when one of the observations is left out.

The general solution is to model \(y_i\) as a missing observation and estimate it
along with all of the model parameters. For a multivariate normal model
\(\log p(y_i\,|\,y_{-i})\) can be computed as follows. First, we model \(y_i\) as
missing and denote the corresponding \emph{parameter} \(y_i^{\mathrm{mis}}\). Then, we
define
\begin{equation}
y_{\mathrm{mis}(i)} = (y_1, \ldots, y_{i-1}, y_i^{\mathrm{mis}}, y_{i+1}, \ldots, y_N).
\end{equation}
to be the same as the full set of observations \(y\) but replacing \(y_i\) with
the parameter \(y_i^{\mathrm{mis}}\).
Second, we compute the log predictive densities as in Equations \eqref{lp-normal}
and \eqref{lp-student}, but replacing \(y\) with
\(y_{\mathrm{mis}(i)}\) in all computations.
Finally, the leave-one-out predictive distribution can be estimated as
\begin{equation}
 p(y_i\,|\,y_{-i}) \approx \frac{1}{S} \sum_{s=1}^S p(y_i\,|\,y_{-i}, \theta_{-i}^{(s)}),
\end{equation}
where \(\theta_{-i}^{(s)}\) are draws from the posterior distribution
\(p(\theta\,|\,y_{\mathrm{mis}(i)})\).

\hypertarget{case-study}{%
\section{Case Study: Neighborhood Crime in Columbus, Ohio}\label{case-study}}

In order to demonstrate how to carry out the computations implied by
these equations, we will fit and evaluate lagged SAR models to data on crime
in 49 different neighborhoods of Columbus, Ohio during the year 1980.
The data was originally described in \citep{anselin1988} and ships with the
spdep R package \citep{bivand2015}.
The three variables in the data set relevant to this example are:
\texttt{CRIME}: the number of residential burglaries and vehicle thefts per
thousand households in the neighborhood, \texttt{HOVAL}: housing value in units
of \$1000 USD, and \texttt{INC}: household income in units of \$1000 USD. In
addition, we have information about the spatial relationship of neighborhoods
from which we can construct the weight matrix to help account for the spatial
dependency among the observations. In addition to the loo R package \citep{loo2018},
for this analysis we use the brms interface \citep{brms1, brms2} to Stan
\citep{carpenter2017} to generate a Stan program and fit the model. The complete R
code for this case study can be found in the supplemental materials.

\begin{figure}
\centering
\includegraphics{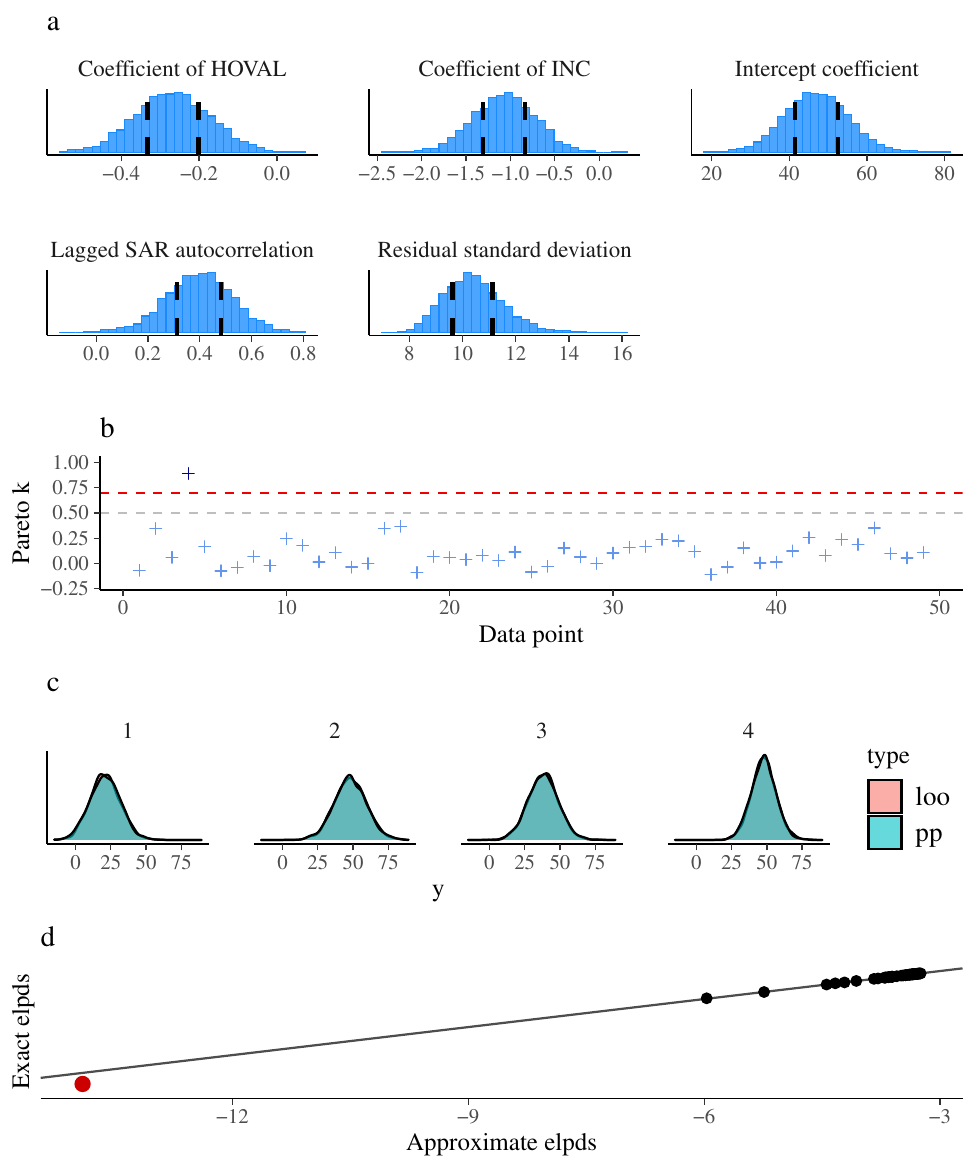}
\caption{\label{fig:plots-normal}Results of the normal SAR model. 1) Posterior distribution of selected parameters of the lagged SAR model along with posterior median and 50\% central interval. 2) PSIS diagnostic plot showing the Pareto-\(k\)-estimate of each observation. 3) Implied response values of the first four observations computed (a) after model fitting (type = `loo') and (b) as part of the model in the form of posterior-predictive draws for the missing observation (type = `pp'). As both distributions are almost identical, the `loo' distribution is hidden behind the `pp' distribution. 4) Comparison of approximate and exact pointwise elpd values. Problematic observations are marked as red dots.}
\end{figure}

We fit a normal SAR model first using the weakly-informative default priors
of brms. In Figure \ref{fig:plots-normal} (a), we see that
both higher income and higher housing value predict lower crime rates in the
neighborhood. Moreover, there seems to be substantial spatial correlation
between adjacent neighborhoods, as indicated by the posterior distribution of
the \texttt{lagsar} parameter.

In order to evaluate model fit, the next step is to compute the pointwise
log-likelihood values needed for approximate LOO-CV and we apply the
method laid out in Section \ref{approx-loo-cv}. Since this is already implemented in
brms\footnote{Source code is available at
  \url{https://github.com/paul-buerkner/brms/blob/master/R/log_lik.R}.}, we can simply use the built-in \texttt{loo} method on the fitted model to obtain the desired results.
The quality of the approximation can be investigated graphically by plotting the Pareto-\(k\) diagnostic for each observation. Ideally, they should not exceed \(0.5\), but in practice the algorithm turns out to be robust up to values of \(0.7\) \citep{vehtari2017loo, vehtari2019psis}. In Figure \ref{fig:plots-normal} (b), we see that the fourth observation is problematic.
This has two implications. First, it may reduce the accuracy of the LOO-CV approximation.
Second, it indicates that the fourth observation is highly influential for the
posterior and thus questions the robustness of the inference obtained by means
of this model. We will address the former issue first and come back to the
latter issue afterwards.

The PSIS-LOO-CV approximation of the expected log predictive density for new
data reveals \(\text{elpd}_{\text{approx}} =\) -186.9.
To verify the correctness of our approximate estimates, this result still needs to be
validated against exact LOO-CV, which is somewhat more involved, as we need to
re-fit the model \(N\) times each time leaving out a single observation. For the
lagged SAR model, we cannot just ignore the held-out observation entirely as
this will change the prior distribution. In other words, the lagged SAR model
does not have the marginalization property that holds, for instance, for
Gaussian process models. Instead, we have to model the held-out observation as a
missing value, which is to be estimated along with the other model parameters
(see the supplemental material for details on the R code).

A first step in the validation of the pointwise predictive density is to compare
the distribution of the implied response values for the left-out observation
using the pointwise mean and standard deviation from (see Proposition
\ref{prop-eff-normal}) to the distribution of the \(y_i^{\mathrm{mis}}\)
posterior-predictive values estimated as part of the model. If the pointwise
predictive density is correct, the two distributions should match very closely
(up to sampling error). In Figure \ref{fig:plots-normal} (c), we overlay these two
distributions for the first four observations and see that they match very
closely (as is the case for all \(49\) observations in this example).

In the final step, we compute the pointwise predictive density based on the
exact LOO-CV and compare it to the approximate PSIS-LOO-CV result computed
earlier. The results of the approximate (\(\text{elpd}_{\text{approx}} =\)
-186.9) and exact LOO-CV (\(\text{elpd}_{\text{exact}} =\) -188.1)
are similar but not as close as we would expect if there were no
problematic observations. We can investigate this issue more closely by plotting
the approximate against the exact pointwise ELPD values.
In Figure \ref{fig:plots-normal} (d), the fourth data point -- the observation
flagged as problematic by the PSIS-LOO approximation -- is colored in red and is
the clear outlier. Otherwise, the correspondence between the exact and
approximate values is strong. In fact, summing over the pointwise ELPD values
and leaving out the fourth observation yields practically equivalent results for
approximate and exact LOO-CV (\(\text{elpd}_{\text{approx},-4} =\)
-173.0 vs.
\(\text{elpd}_{\text{exact},-4} =\) -173.0).
From this we can conclude that the difference we
found when including \emph{all} observations does not indicate an error in the
implementation of the approximate LOO-CV but rather a violation of its
assumptions.

With the correctness of the approximating procedure established for
non-problematic observations, we can now go ahead and correct for the problematic
observation in the approximate LOO-CV estimate. \citet{vehtari2017loo} recommend to
perform exact LOO-CV only for the problematic observations and replace their
approximate ELPD contributions with their exact counterparts \citep[see also][ for an alternative method]{paananen2019}. So this time, we do not use exact LOO-CV for validation of
the approximation but rather to improve the latter's accuracy when needed. In
the present normal SAR model, only the 4th observation was diagnosed as
problematic and so we only need to update the ELPD contribution of this
observation. The results of the corrected approximate
(\(\text{elpd}_{\text{approx}} =\) -188.0) and
exact LOO-CV (\(\text{elpd}_{\text{exact}} =\) -188.1) are now
almost equal for the complete data set as well.

\begin{figure}
\centering
\includegraphics{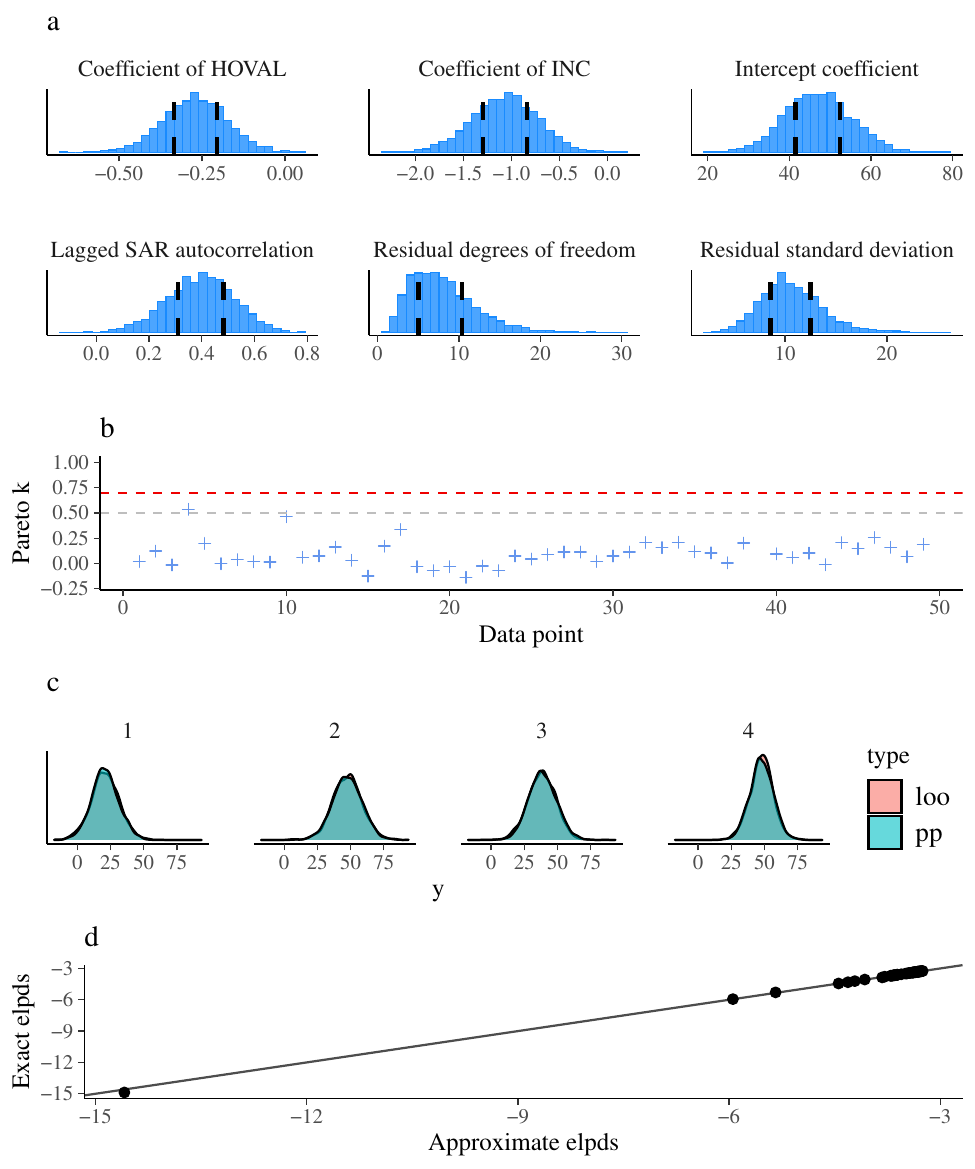}
\caption{\label{fig:plots-student}Results of the Student-\(t\) SAR model. a) Posterior distribution of selected parameters of the lagged SAR model along with posterior median and 50\% central interval. b) PSIS diagnostic plot showing the Pareto-\(k\)-estimate of each observation. c) Implied response values of the first four observations computed (1) after model fitting (type = `loo') and (2) as part of the model in the form of posterior-predictive draws for the missing observation (type = `pp'). As both distributions are almost identical, the `loo' distribution is hidden behind the `pp' distribution. d) Comparison of approximate and exact pointwise elpd values. There were no problematic observations for this model.}
\end{figure}

Although we were able to correct for the problematic observation in the
approximate LOO-CV estimation, the mere existence of such problematic
observations raises doubts about the appropriateness of the normal SAR model for
the present data. Accordingly, it is sensible to fit a Student-\(t\) SAR model as a
potentially better predicting alternative due to its fatter tails. We choose an
informative \({\rm Gamma}(4, 0.5)\) prior (with mean \(8\) and standard deviation
\(4\)) on the degrees of freedom parameter \(\nu\) to ensure rather fat tails of the
likelihood a-priori. For all other parameters, we continue to use the
weakly-informative default priors of brms. In Figure \ref{fig:plots-student}
(a), the marginal posterior distributions of the main model parameters are
depicted. Comparing the results to those shown in Figure \ref{fig:plots-normal}
(a), we see that the estimates of both the regression parameters and the SAR
autocorrelation are quite similar to the estimates obtained from the normal
model.

In contrast to the normal case, we see in Figure \ref{fig:plots-student} (b) that the 4th observation is no longer recognized as problematic by the Pareto-\(k\) diagnostics.
It does exceed \(0.5\) slightly but does not exceed the more important threshold
of \(0.7\) above which we would stop trusting the PSIS-LOO-CV approximation.
Indeed, comparison between the approximate (\(\text{elpd}_{\text{approx}} =\)
-187.7) and exact LOO-CV
(\(\text{elpd}_{\text{exact}} =\) -187.9) based on the complete
data demonstrates that they are very similar (up to random error due to the MCMC
estimation). The results shown in Figure \ref{fig:plots-student} (c) and (d)
have the same interpretation as the analogous plots for the normal case and
provide further evidence for both the correctness of our (exact and
approximate) LOO-CV methods for non-factorized Student-\(t\) models and for the
quality of the PSIS-LOO-CV approximation for the present Student-\(t\) SAR model.

\begin{figure}
\centering
\includegraphics{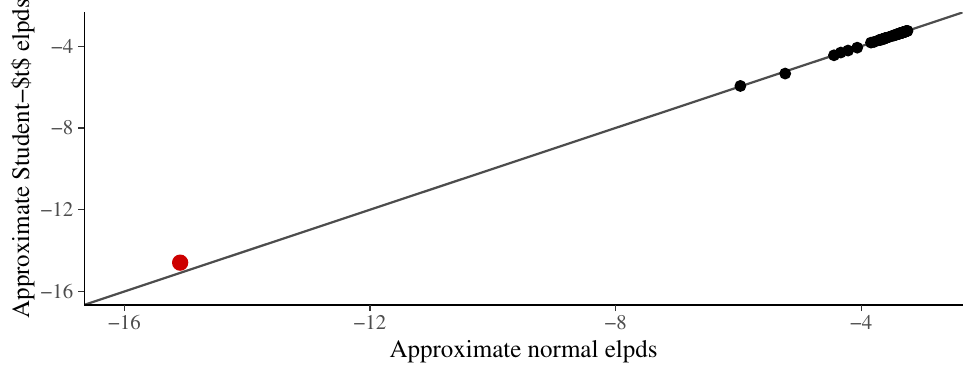}
\caption{\label{fig:elpd-approx-compare}Comparison of approximate pointwise elpd values for the normal SAR model (after refit for the 4th observation) and the Student-\(t\) SAR model (without refit). Observations with relevant differences are highlighted in red.}
\end{figure}

Lastly, let us compare the PSIS-LOO-CV estimate of the normal SAR model (after
correcting for the problematic observation via refit) to the Student-\(t\) SAR
model. The ELPD difference between the two models is -0.3
(SE = 0.5) in favor of the Student-\(t\) model, and thus very
small and not substantial for any practical purposes. As shown in Figure
\ref{fig:elpd-approx-compare}, the pointwise elpd contributions are also highly
similar. The Student-\(t\) model fits slightly but noticeably better only for the 4th
observation. Other methods clearly identify the 4th observation as problematic
as well \citep[e.g.,][]{halleck2015}. However, what exactly causes this outlier remains
unclear so far as nobody has been able to verify or spatially register the data
set despite many attempts.

\hypertarget{conclusion}{%
\section{Conclusion}\label{conclusion}}

In this paper we derived how to perform and validate exact and approximate
leave-one-out cross-validation (LOO-CV) for non-factorized multivariate normal
and Student-\(t\) models and we demonstrated the practical applicability of our
method in a case study using spatial autoregressive models. The proposed LOO-CV
approximation makes efficient and robust model fit evaluation and comparison
feasible for many models that are widely used in temporal and spatial
statistics. Importantly, our method does not only apply to non-factorizable
models for which a factorized likelihood is unavailable. It is also useful when
factorization is possible in principle but could result in a unstable LOO
predictive density, either for numerical reasons or due to the use of finite
estimation procedures like Markov-Chain Monte Carlo. In such cases,
marginalizing over observation-specific latent variables and using a
non-factorized formulation can lead to more robust estimates.

\hypertarget{acknowledgements}{%
\section{Acknowledgements}\label{acknowledgements}}

We thank Daniel Simpson for useful discussion and the Academy of Finland (grants
298742, 313122) as well as the Technology Industries of Finland Centennial
Foundation (grant 70007503; Artificial Intelligence for Research and
Development) for partial support of this work.

\hypertarget{appendix}{%
\section*{Appendix}\label{appendix}}
\addcontentsline{toc}{section}{Appendix}

\emph{Proof} of Proposition \ref{prop-eff-normal}.
In their Lemma 1, \citet{sundararajan2001} prove for any finite subset \(z\) of a
zero-mean Gaussian process with covariance matrix \(\Sigma\) that the LOO predictive
mean and standard deviation can be computed as
\begin{equation}
  \tilde{\mu}_{i} = z_i - \frac{g_i}{\bar{\sigma}_{ii}},
\end{equation}
\begin{equation}
  \tilde{\sigma}_{i} = \frac{1}{\bar{\sigma}_{ii}},
\end{equation}
where \(g_i = \left[\Sigma^{-1} z\right]_i\) and \(\bar{\sigma}_{ii} = \left[\Sigma^{-1}\right]_{ii}\). Their proof does not make use of any specific
form of \(\Sigma\) and thus directly applies to all zero-mean multivariate normal
distributions. If \(y\) is multivariate normal with mean \(\mu\) then \((y - \mu)\) is
multivariate normal with mean \(0\) and unchanged covariance matrix.
Thus, we can replace \(z\) with \((y - \mu)\) in
the above equations. By the same argument we see that, if \((y_i - \mu_i)\) has LOO
mean \((y_i - \mu_i) - \frac{g_i}{\bar{\sigma}_{ii}}\), then \(y\) has LOO mean \(y_i - \frac{g_i}{\bar{\sigma}_{ii}}\) which completes the proof. \hfill \(\Box\)

\emph{Proof} of Proposition \ref{prop-cond-student}.
Using the parameterization \(K := {\rm Cov}(y) = \frac{\nu}{\nu - 2} \Sigma\) and
requiring \(\nu > 2\), \citet{shah2014} proof in their Lemma 3 that, if \(y = (y_1, y_2)\)
is multivariate Student-\(t\) of dimension \(N = N_1 + N_2\), then \(y_2\) given \(y_1\)
is multivariate Student-\(t\) of dimension \(N_2\). Moreover, they provide equations
for the parameters of the conditional Student-\(t\) distribution. When we
parameterize for \(\Sigma\) instead of \(K\) and allow for \(\nu > 0\), we can repeat
their proof analogously which yields the following parameters of the conditional
Student-\(t\) distribution of \(y_2\) given \(y_1\):
\begin{equation}
\tilde{\nu}_2 = \nu + N_1,
\end{equation}
\begin{equation}
  \tilde{\mu}_{2} = \mu_2 + \Sigma_{2,1} \Sigma^{-1}_{1}(y_1 - \mu_{1}),
\end{equation}
\begin{equation}
  \tilde{\sigma}_{2} = \frac{\nu + \beta_{1}}{\nu + N_1} 
  \left( \Sigma_{22} + \Sigma_{2,1} \Sigma^{-1}_{1} \Sigma_{1,2} \right),
\end{equation}
with
\begin{equation}
\beta_{1} = (y_{1} - \mu_{1})^{\rm T} \Sigma^{-1}_{1} (y_{1} - \mu_{1}).
\end{equation}
where we use the subscripts \(1\) and \(2\) to refer to the \(1\)st and \(2\)nd subset
of \(y\), respectively. Setting \(y_1 = y_{-i}\), \(y_2 = y_i\) for \(i = 1, \ldots, N\)
and noting that \(N_1 = N_{-i} = N - 1\) completes the proof.
\hfill \(\Box\)

\emph{Proof} of Proposition \ref{prop-eff-student}.
The correctness of Equations \eqref{cmeant2} and \eqref{csdt2} follows directly
from Equations \eqref{cmean2}, and \eqref{csd2}. To show \eqref{cbeta2}, we
perform a rank-one update of \(\Sigma^{-1}\) as per Theorem 2.1 of \citet{juarez2016} based on the Sherman-Morrison formula \citep{bartlett1951, sherman1950}. In general, if we exclude row
\(p\) and column \(q\) from \(\Sigma\), the inverse \(\Sigma^{-1}_{-p,-q}\) of \(\Sigma_{-p,-q}\) exists if \(\sigma_{pq} \neq 0\) and \(\bar{\sigma}_{pq} \neq 0\). The
elements \(m_{jk}\) (\(j,k = 1,\ldots,N\), \(j \neq p\), \(k \neq q\)) of \(\Sigma^{-1}_{-p,-q}\)
are then given by
\begin{equation}
m_{jk} = \bar{\sigma}_{jk} - 
  \frac{\bar{\sigma}_{jp} \bar{\sigma}_{qk}}{\bar{\sigma}_{pq}}.
\end{equation}
where \(\bar{\sigma}_{jk}\) is the \((j,k)\)th element of \(\Sigma^{-1}\).
We now set \(p = q = i\) and note that \(\sigma_{ii} > 0\) and \(\bar{\sigma}_{ii} > 0\)
since \(\Sigma\) is a covariance matrix, which leads to
\begin{equation}
m_{jk} = \bar{\sigma}_{jk} -
  \frac{\bar{\sigma}_{ji} \bar{\sigma}_{ik}}{\bar{\sigma}_{ii}}.
\end{equation}
for each \(i = 1,\ldots,N\). Switching to matrix notation completes the proof.
\hfill \(\Box\)

  \bibliography{psis-non-factorized-models}

\begin{thebibliography}{}

\bibitem[Ando and Tsay, 2010]{ando2010}
Ando, T. and Tsay, R. (2010).
\newblock Predictive likelihood for {Bayesian} model selection and averaging.
\newblock {\em International Journal of Forecasting}, 26(4):744--763.

\bibitem[Anselin, 1988]{anselin1988}
Anselin, L. (1988).
\newblock {\em Spatial econometrics: methods and models}.
\newblock Dordrecht: Kluwer Academic.

\bibitem[Bartlett, 1951]{bartlett1951}
Bartlett, M.~S. (1951).
\newblock An inverse matrix adjustment arising in discriminant analysis.
\newblock {\em The Annals of Mathematical Statistics}, 22(1):107--111.

\bibitem[Bivand and Piras, 2015]{bivand2015}
Bivand, R. and Piras, G. (2015).
\newblock Comparing implementations of estimation methods for spatial
  econometrics.
\newblock {\em Journal of Statistical Software}, 63(18):1--36.

\bibitem[B{\"u}rkner, 2017]{brms1}
B{\"u}rkner, P.-C. (2017).
\newblock {brms}: An {R} package for {Bayesian} multilevel models using {Stan}.
\newblock {\em Journal of Statistical Software}, 80(1):1--28.

\bibitem[B{\"u}rkner, 2018]{brms2}
B{\"u}rkner, P.-C. (2018).
\newblock Advanced {Bayesian} multilevel modeling with the {R} package {brms}.
\newblock {\em The R Journal}, pages 395--411.

\bibitem[Carpenter et~al., 2017]{carpenter2017}
Carpenter, B., Gelman, A., Hoffman, M., Lee, D., Goodrich, B., Betancourt, M.,
  Brubaker, M.~A., Guo, J., Li, P., and Ridell, A. (2017).
\newblock Stan: A probabilistic programming language.
\newblock {\em Journal of Statistical Software}.

\bibitem[Cressie, 1992]{cressie1992}
Cressie, N. (1992).
\newblock Statistics for spatial data.
\newblock {\em Terra Nova}, 4(5):613--617.

\bibitem[Fern{\'a}ndez and Steel, 1999]{fernandez1999}
Fern{\'a}ndez, C. and Steel, M.~F. (1999).
\newblock Multivariate student-t regression models: Pitfalls and inference.
\newblock {\em Biometrika}, 86(1):153--167.

\bibitem[Geisser and Eddy, 1979]{geisser1979}
Geisser, S. and Eddy, W.~F. (1979).
\newblock A predictive approach to model selection.
\newblock {\em Journal of the American Statistical Association},
  74(365):153--160.

\bibitem[Gelfand et~al., 1992]{gelfand1992}
Gelfand, A., Dey, D., and Chang, H. (1992).
\newblock Model determination using predictive distributions with
  implementation via sampling-based methods.
\newblock {\em Bayesian Statistics}, 4:147--167.

\bibitem[Gelfand and Vounatsou, 2003]{gelfand2003}
Gelfand, A.~E. and Vounatsou, P. (2003).
\newblock Proper multivariate conditional autoregressive models for spatial
  data analysis.
\newblock {\em Biostatistics}, 4(1):11--15.

\bibitem[Goulard et~al., 2017]{goulard2017}
Goulard, M., Laurent, T., and Thomas-Agnan, C. (2017).
\newblock About predictions in spatial autoregressive models: Optimal and
  almost optimal strategies.
\newblock {\em Spatial Economic Analysis}, 12(2-3):304--325.

\bibitem[Haining and Haining, 2003]{haining2003}
Haining, R.~P. and Haining, R. (2003).
\newblock {\em Spatial data analysis: theory and practice}.
\newblock Cambridge university press.

\bibitem[Halleck~Vega and Elhorst, 2015]{halleck2015}
Halleck~Vega, S. and Elhorst, J.~P. (2015).
\newblock The slx model.
\newblock {\em Journal of Regional Science}, 55(3):339--363.

\bibitem[Hoeting et~al., 1999]{hoeting1999}
Hoeting, J.~A., Madigan, D., Raftery, A.~E., and Volinsky, C.~T. (1999).
\newblock Bayesian model averaging: a tutorial.
\newblock {\em Statist. Sci.}, 14(4):382--417.

\bibitem[Ju{\'a}rez-Ruiz et~al., 2016]{juarez2016}
Ju{\'a}rez-Ruiz, E., Cort{\'e}s-Maldonado, R., and P{\'e}rez-Rodr{\'\i}guez, F.
  (2016).
\newblock Relationship between the inverses of a matrix and a submatrix.
\newblock {\em Computaci{\'o}n y Sistemas}, 20(2):251--262.

\bibitem[LeSage and Pace, 2009]{lesage2009}
LeSage, J.~P. and Pace, R.~K. (2009).
\newblock {\em Introduction to Spatial Econometrics}.
\newblock CRC Press.

\bibitem[O'Hagan, 1979]{ohagan1979}
O'Hagan, A. (1979).
\newblock On outlier rejection phenomena in {Bayes} inference.
\newblock {\em Journal of the Royal Statistical Society: Series B
  (Methodological)}, 41(3):358--367.

\bibitem[Paananen et~al., 2019]{paananen2019}
Paananen, T., Piironen, J., B{\"u}rkner, P.-C., and Vehtari, A. (2019).
\newblock Pushing the limits of importance sampling through iterative moment
  matching.
\newblock {\em arXiv preprint arXiv:1906.08850}.

\bibitem[Pich{\'e} et~al., 2012]{piche2012}
Pich{\'e}, R., S{\"a}rkk{\"a}, S., and Hartikainen, J. (2012).
\newblock Recursive outlier-robust filtering and smoothing for nonlinear
  systems using the multivariate student-t distribution.
\newblock In {\em 2012 IEEE International Workshop on Machine Learning for
  Signal Processing}, pages 1--6. IEEE.

\bibitem[Rabiner and Juang, 1986]{rabiner1986}
Rabiner, L. and Juang, B. (1986).
\newblock An introduction to hidden {Markov} models.
\newblock {\em ieee assp magazine}, 3(1):4--16.

\bibitem[Rasmussen, 2003]{rasmussen2003}
Rasmussen, C.~E. (2003).
\newblock Gaussian processes in machine learning.
\newblock In {\em Summer School on Machine Learning}, pages 63--71. Springer.

\bibitem[Shah et~al., 2014]{shah2014}
Shah, A., Wilson, A., and Ghahramani, Z. (2014).
\newblock Student-t processes as alternatives to {Gaussian} processes.
\newblock In {\em Artificial intelligence and statistics}, pages 877--885.

\bibitem[Sherman and Morrison, 1950]{sherman1950}
Sherman, J. and Morrison, W.~J. (1950).
\newblock Adjustment of an inverse matrix corresponding to a change in one
  element of a given matrix.
\newblock {\em The Annals of Mathematical Statistics}, 21(1):124--127.

\bibitem[Sundararajan and Keerthi, 2001]{sundararajan2001}
Sundararajan, S. and Keerthi, S.~S. (2001).
\newblock Predictive approaches for choosing hyperparameters in {Gaussian}
  processes.
\newblock {\em Neural Computation}, 13(5):1103--1118.

\bibitem[Tong, 2012]{tong2012}
Tong, Y.~L. (2012).
\newblock {\em The multivariate normal distribution}.
\newblock Springer Science \& Business Media.

\bibitem[Vehtari et~al., 2018]{loo2018}
Vehtari, A., Gabry, J., Yao, Y., and Gelman, A. (2018).
\newblock {\em {\bf loo}: {E}fficient Leave-One-Out Cross-Validation and {WAIC}
  for {B}ayesian Models.}
\newblock R package version 2.0.0.

\bibitem[Vehtari et~al., 2017]{vehtari2017loo}
Vehtari, A., Gelman, A., and Gabry, J. (2017).
\newblock Practical {Bayesian} model evaluation using leave-one-out
  cross-validation and {WAIC}.
\newblock {\em Statistics and Computing}, 27(5):1413--1432.

\bibitem[Vehtari and Ojanen, 2012]{vehtari2012}
Vehtari, A. and Ojanen, J. (2012).
\newblock A survey of {Bayesian} predictive methods for model assessment,
  selection and comparison.
\newblock {\em Statistics Surveys}, 6:142--228.

\bibitem[Vehtari et~al., 2019]{vehtari2019psis}
Vehtari, A., Simpson, D., Gelman, A., Yao, Y., and Gabry, J. (2019).
\newblock Pareto smoothed importance sampling.
\newblock {\em arXiv preprint}.

\bibitem[Welch et~al., 1995]{welch1995}
Welch, G., Bishop, G., et~al. (1995).
\newblock An introduction to the {Kalman} filter.

\bibitem[Zellner, 1976]{zellner1976}
Zellner, A. (1976).
\newblock Bayesian and non-{Bayesian} analysis of the regression model with
  multivariate {Student}-t error terms.
\newblock {\em Journal of the American Statistical Association},
  71(354):400--405.

\bibitem[Zhang and Yeung, 2010]{zhang2010}
Zhang, Y. and Yeung, D.-Y. (2010).
\newblock Multi-task learning using generalized t process.
\newblock In {\em Proceedings of the Thirteenth International Conference on
  Artificial Intelligence and Statistics}, pages 964--971.

\end{thebibliography}

\end{document}